\documentclass{eas}
\usepackage{graphicx}

\begin{document}

\TitreGlobal{Mass Profiles and Shapes of Cosmological Structures}

\title{The Draco dwarf in CDM and MOND}
\author{{\L}okas, E.}\address{Nicolaus Copernicus Astronomical Center, Warsaw, Poland}
\author{Mamon, G.}\address{Institut d'Astrophysique de Paris, France}
\author{Prada, F.}\address{Instituto de Astrof{\'\i}sica de Andalucia, Granada, Spain}
\runningtitle{The Draco dwarf in CDM and MOND}
\setcounter{page}{1}
\index{{\L}okas, E.}
\index{Mamon, G.}
\index{Prada, F.}

\maketitle
\begin{abstract}
We present interpretations of the line-of-sight velocity distribution of stars in the
Draco dwarf spheroidal galaxy in terms of CDM and MOND assuming constant mass-to-light ratio
and anisotropy. We estimate the two parameters by fitting
the line-of-sight velocity dispersion and kurtosis profiles for stellar samples
differing by a number of stars rejected as interlopers.
The results of the fitting procedure for CDM, high mass-to-light ratio
(131-141 solar units in V-band) and weakly tangential orbits, are similar for
different samples, but the quality of the fit is much worse when fewer interlopers
are removed. For MOND, the derived mass-to-light ratio (21 solar units) is too large to be
explained by the stellar content of the galaxy.
\end{abstract}

\section{Introduction}

The Draco dwarf is a generic example of the class of dwarf spheroidal (dSph) galaxies of the Local Group
and as such has been a subject of intensive study in recent years. The object is interesting
from the point of view of theories of structure formation due to its large dark matter content
and its implications for the missing satellites problem. Dwarf spheroidals are also in the regime
of low accelerations and therefore can provide critical tests for the alternatives to cold
dark matter (CDM) scenarios, such as the Modified Newtonian Dynamics (MOND). New observations of
the object performed recently provided a larger kinematic sample of stellar velocities
(Kleyna et al. 2002; Wilkinson et al. 2004) and a better determination of its luminosity
profile and shape (Odenkirchen et al. 2001).

This has encouraged us to study the Draco dwarf using a method of velocity moments
({\L}okas, Mamon \&
Prada 2005) first applied to the Coma cluster of galaxies by {\L}okas \& Mamon (2003) and
tested with $N$-body simulations by Sanchis, {\L}okas \& Mamon (2004) and Wojtak et al. (2005).
The method relies on joint fitting
the velocity dispersion and kurtosis profiles which allows us to break the degeneracy between
the mass distribution and velocity anisotropy. We have assumed that the dark matter density profile
is given by a formula with an inner cusp and exponential cut-off at large distances recently proposed
by Kazantzidis et al. (2004) as a result of their simulations of tidal stripping of dwarfs by
the potential of a Milky Way size galaxy. We found that the results depend dramatically on the sample
of stars under consideration and concluded that the larger samples are probably contaminated by
unbound stars because the velocity moments constructed from them are inconsistent.

Since this dark matter profile is not very different in shape from the distribution of stars given
by the Sersic profile, in this contribution we reanalyze the same data assuming a simpler
hypothesis that mass traces light and using different binning. This case is directly comparable to
the alternative interpretation of the data in terms of MOND (Sanders \& McGaugh 2004)
which also assumes constant mass-to-light
ratio but modifies the gravitational acceleration. We examine the two hypotheses subsequently in the
next two sections.

\section{Cold Dark Matter}

\begin{figure}[t]
   \centering
   \includegraphics[width=12.6cm]{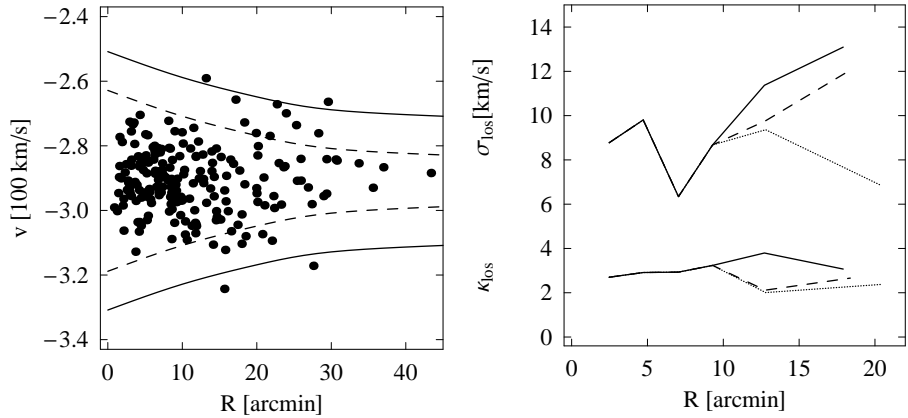}
      \caption{Left panel: line-of-sight velocities versus distances for 207 stars in the sample
	of Wilkinson et al. (2004) with lines separating supposed interlopers from members.
	Right panel: velocity moments calculated from different samples with 0, 4 and 18 interlopers
	removed (solid, dashed and dotted lines respectively).}
      \label{figure1}
\end{figure}

The kinematic data we have used are plotted in the left panel of Fig.~\ref{figure1}
which shows the line-of-sight
velocities versus distances of 207 stars counted as members of Draco by Wilkinson et al. (2004).
Since some of the stars clearly are discrepant from the main body of the galaxy we further exclude
some of them as interlopers. The selection is indicated by the solid and dashed lines intended to
follow the overall shape of such diagrams for gravitationally bound object. The line-of-sight
velocity moments, dispersion $\sigma_{\rm los}$ and kurtosis $\kappa_{\rm los}$,
calculated from the different samples thus obtained are plotted in the right panel of the Figure. In
each case we divide the data into 6 radial bins with 30 stars. We see that the results
in the outer bins from which the
interlopers are removed are affected dramatically and both moments are significantly reduced
in value. We have assumed that the binary fraction in the stellar sample is small and its
effect on the moment negligible in comparison with other uncertainties (but see the
discussion in {\L}okas et al. 2005).

\begin{table}
\caption{Fitted parameters in the case of CDM }
\label{parameters}
\begin{center}
\begin{tabular}{cccccccc}
number of & & fitting $\sigma_{\rm los}$ & & \ \ \  & \ fitting & \ \ $\sigma_{\rm los}$ & and \ \ \
$\kappa_{\rm los}$ \\
interlopers  & $M/L_V$ & $\beta$ & $\chi^2/N$ & & $M/L_V$ & $\beta$ & $\chi^2/N$  \\
\hline
18   & 131 & -0.44 & 11.1/4 & & 131 & -0.34 & 15.4/10 \\
4    & 147 & -1.24 & 21.7/4 & & 141 & -0.50 & 27.6/10 \\
0    & 156 & -1.86 & 25.9/4 & & 141 & -0.19 & 39.1/10 \\
\hline
\end{tabular}
\end{center}
\end{table}

\begin{figure}[t]
   \centering
   \includegraphics[width=12.6cm]{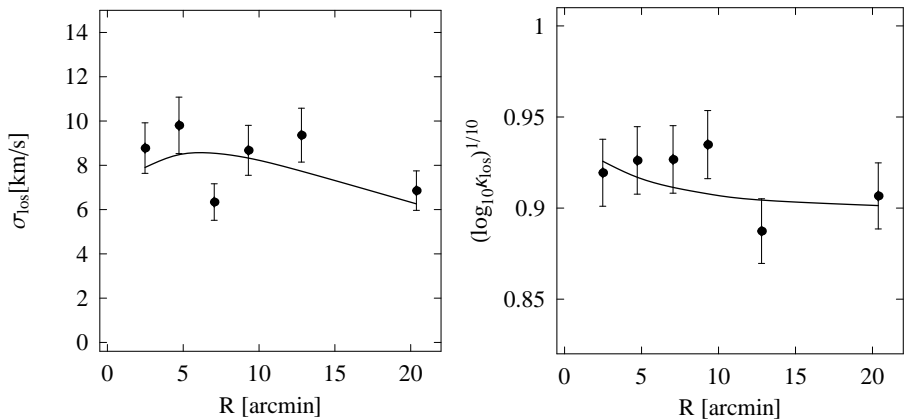}
      \caption{Velocity moments for the most restrictive sample (with 18 interlopers removed)
	together with the best-fitting model in the case of CDM.}
      \label{figure2}
\end{figure}

Assigning sampling errors to the moments we fit them with the models based on
solutions of the Jeans equations characterized by only two
constant parameters, the mass-to-light ratio in V-band, $M/L_V$, and the anisotropy parameter
$\beta=1-\sigma_\theta^2/\sigma_r^2$. The remaining assumptions and adopted parameters are as in
{\L}okas et al. (2005). The results for different samples are listed in Table~\ref{parameters}.

Fitting
first the velocity dispersion alone we find a stronger preference for tangential orbits in the
case of samples with smaller number of interlopers removed. This is understandable: since the shape
of the mass profile is constrained to be strongly decreasing
by the assumption of constant $M/L_V$ the strongly increasing
dispersion profile induces tangential orbits. The tendency disappears, however, once the kurtosis
is included in the analysis; then all samples yield similar best-fitting parameters although the
quality of the fit is much worse for the less restrictive samples. The reason for this is the fact that
for these samples the moments seem inconsistent: if the increasing dispersion profile is due to
tangential orbits then this should result in decreasing instead of increasing kurtosis profile
(see the discussion in {\L}okas et al. 2005).

\begin{figure}[t]
   \centering
   \includegraphics[width=12.6cm]{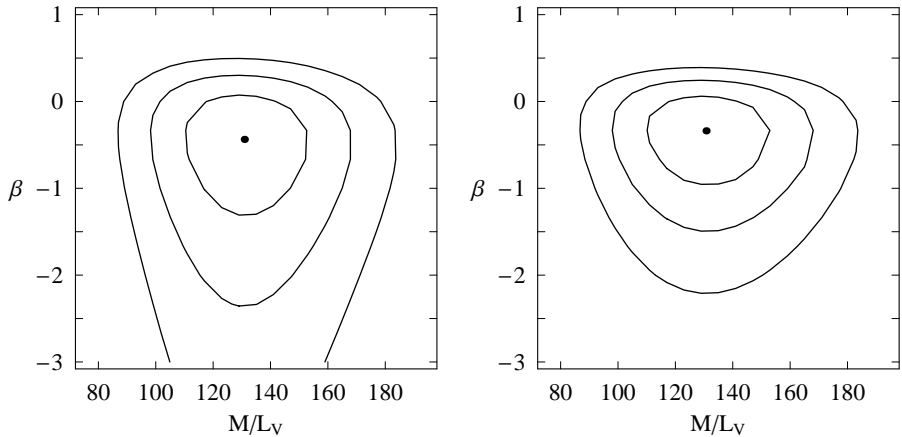}
      \caption{The $1\sigma$, $2\sigma$ and $3\sigma$ confidence regions in the parameter plane
       for the most restrictive sample (with 18 interlopers removed) obtained from fitting velocity
	dispersion alone (left panel) and both moments (right panel) in the case of CDM.}
      \label{figure3}
\end{figure}

The velocity moments together with the best-fitting models for the case of the most restrictive
sample (with 18 interlopers) are shown in Fig.~\ref{figure2}. Fig.~\ref{figure3} illustrates the
benefit from including the kurtosis in the analysis for the same sample. While the best-fitting
parameters are similar in both cases, the confidence limits for $\beta$ one can read from the contours
shown in the Figure are much more
constrained from fitting both moments (right panel) than from fitting the dispersion alone
(left panel).

The mass-to-light ratio $M/L_V=131 M_{\odot}/L_{\odot}$ found here for the most restrictive sample is
similar to the minimum value of 134 reached at 10 arcmin for the corresponding sample of 189 stars
considered by {\L}okas et al. (2005) (see their Fig. 6). Note however, that the quality of the fit
is somewhat worse here because of a smaller number of parameters.

\section{Modified Newtonian Dynamics}

The lowest order Jeans equation which models the velocity dispersion can be simply modified in the
case of MOND by replacing the Newtonian gravitational acceleration $g_{\rm N}$ with
$g_{\rm M}$ related by $g_{\rm N} = \mu(g_{\rm M}/a_0) g_{\rm M}$ where the function $\mu$ is chosen as
$\mu(x)=x(1+x^2)^{-1/2}$ (Milgrom 1983) and $a_0$ is the characteristic acceleration scale of MOND
supposed to be universal. The value found to fit well most rotation curve data for spiral galaxies
is $a_0=1.2 \times 10^{-8}$ cm s$^{-2}$ (Sanders \& McGaugh 2004).

\begin{figure}[t]
   \centering
   \includegraphics[width=12.6cm]{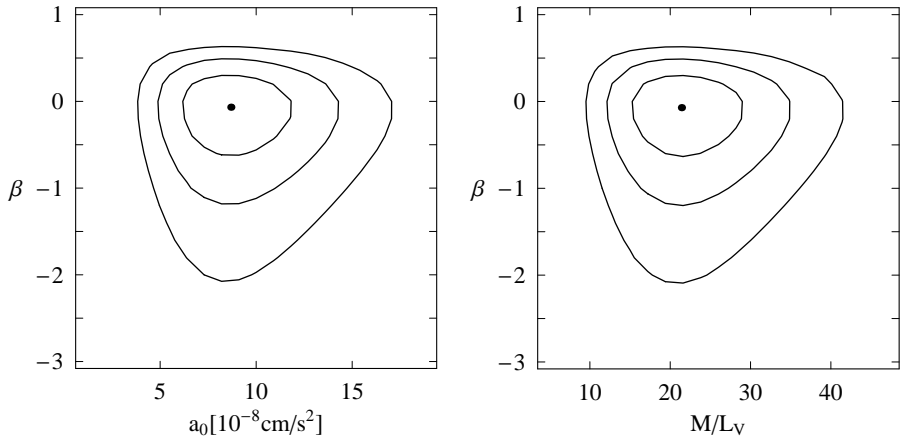}
      \caption{The $1\sigma$, $2\sigma$ and $3\sigma$ confidence regions in the parameter plane
       for the most restrictive sample (with 18 interlopers removed) obtained from fitting velocity
	dispersion in the case of MOND assuming standard $M/L_V$ (left panel) and $a_0$
	(right panel).}
      \label{figure4}
\end{figure}

We have fitted the velocity dispersion data for the most restrictive sample, first assuming
$M/L_V = 3 $M$_{\odot}/$L$_{\odot}$, a value characteristic of the Draco stellar sample, and
finding the best-fitting parameters $a_0$ and $\beta$. Next, we adopted the canonical value
$a_0=1.2 \times 10^{-8}$ cm s$^{-2}$ and fitted $M/L_V$ and $\beta$. The results, in terms of
best-fitting parameters and confidence regions are shown in Fig.~\ref{figure4}. The quality of
the fit, $\chi^2/N=9.1/4$, is the same for both cases.

In agreement
with the earlier findings of {\L}okas (2001, 2002) using older kinematic data, we conclude that
the velocity dispersion profile of Draco cannot be well reproduced with reasonable values of
either $a_0$ or $M/L_V$. The best-fitting $a_0=8.7 \times 10^{-8}$ cm s$^{-2}$ and
$M/L_V = 21.4 \ $M$_{\odot}/$L$_{\odot}$ are about an order of magnitude larger than expected and
the confidence regions are rather narrow excluding the standard values at more than 3$\sigma$
level. These values are even higher (while orbits more tangential and quality of the fits worse)
for the other samples with less interlopers removed.

\section{Conclusion}

We conclude from this study that the constant mass-to-light ratio models provide a reasonably
good fit (with $\chi^2/N=15.4/10$) to the
kinematic data for Draco, but only for the most restrictive sample (with 18 interlopers removed).
For other samples, although similar best-fitting parameters are found from the joint fitting
of both moments, the fits are much worse ($\chi^2/N=27.6/10$ and $\chi^2/N=39.1/10$
respectively for samples with 4 and 0 interlopers removed). This result is in qualitative agreement
with the conclusion reached by {\L}okas et al. (2005) who considered a wider class of dark matter
distributions.

The velocity dispersion data are also poorly fitted by MOND. The best-fitting parameters for
the most restrictive sample, either
$a_0$ or $M/L_V$ are found to be about an order of magnitude larger than expected. However,
before we conclude
that the case of Draco really falsifies MOND other possibilities have to be considered. The present
results might mean that $a_0$ is not really a universal constant in MOND as previously claimed
and takes different values for different classes of objects. Another possibility to be explored is
that in modifying the gravitational potential according to MOND, the influence of the Milky Way
(here neglected) has to be taken into account. In addition, since in MOND the mass is proportional
to the fourth power of velocity dispersion any interlopers still present in our sample may
artificially inflate the mass more strongly than in the case of CDM.

The most interesting issue to address in the future research on dSph galaxies is the determination
of their dark matter content without the uncertainties due to unbound stars.
This could be done either by modelling more distant dwarfs which are not under direct influence of
the giant galaxies of the Local Group or by careful removal of interlopers aided perhaps by
simulations of their origin.


\begin{thebibliography}{}
\bibitem[]{kmmds} Kazantzidis, S., Mayer, L., Mastropietro, C., Diemand, J., Stadel, J.,
	\& Moore, B. 2004, ApJ, 608, 663
\bibitem[]{kl1} Kleyna, J. T., Wilkinson, M. I., Evans, N. W., \& Gilmore, G. 2002,
	MNRAS, 330, 792
\bibitem[]{lok} \L okas, E. L. 2001, MNRAS,  327, 21P
\bibitem[]{lo} {\L}okas, E. L. 2002, MNRAS, 333, 697
\bibitem[]{}{\L}okas, E. L., \& Mamon, G. A. 2003, MNRAS, 343, 401
\bibitem[]{}{\L}okas, E. L., \& Mamon, G. A., \& Prada, F. 2005, MNRAS, in press, astro-ph/0411694
\bibitem[]{mil83b} Milgrom, M. 1983, ApJ, 270, 371
\bibitem[]{oden} Odenkirchen, M., et al. 2001, AJ, 122, 2538
\bibitem[]{}Sanchis, T., {\L}okas, E. L., \& Mamon, G. A. 2004, MNRAS, 347, 1198
\bibitem[]{}Sanders, R. H., \& McGaugh, S. S. 2004, ARA\&A, 40, 263
\bibitem[]{wkeg} Wilkinson, M. I., Kleyna, J. T., Evans, N. W., Gilmore, G. F.,
	Irwin, M. J., \& Grebel, E. K. 2004, ApJ, 611, L21
\bibitem[]{}Wojtak, R., {\L}okas, E. L., Gottl\"ober, S., \& Mamon, G. A. 2005, these proceedings,
	astro-ph/0508639

\end{thebibliography}
\end{document}